# Long-Term Aging Study of a Silicon Nitride Nanomechanical Resonator

Michel Stephan, Alexandre Bouchard, Timothy Hodges, Richard G. Green, Triantafillos Koukoulas, Lixue Wu, Raphael St-Gelais

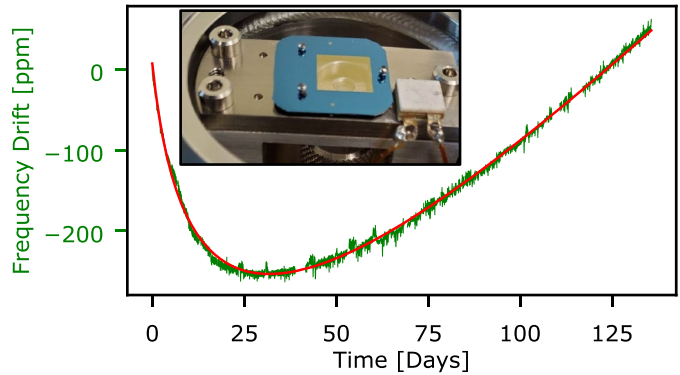

**ABSTRACT** - Short-term changes in the resonance frequency of silicon nitride (SiN) nanomechanical resonators can be measured very precisely due to low thermomechanical fluctuations resulting from large mechanical quality factors. These properties enable high-performance detection of quasi-instantaneous stimuli, such as sudden exposure to radiation or adsorption of mass. However, practical use of such sensors will eventually raise questions regarding their less-studied longer-term stability, notably for calibration purposes. We characterize aging of an as-fabricated SiN membrane by continuously tracking changes of its resonance frequency over 135 days in a temperature-controlled high vacuum environment. The aging behavior is consistent with previously reported double-logarithmic and drift-reversal aging trends observed in quartz oscillators. The aging magnitude (300 ppm) is also comparable to typical temperature compensated quartz oscillators (TCXO), after normalization to account for the greater importance of interfaces in our thin (90 nm) resonator, compared to several microns thick TCXOs. Possible causes of aging due to surface adsorption are investigated. We review models on how water adsorption and desorption can cause significant frequency changes, predominantly due to chemisorption stress. Chemical species adsorbed on the resonator surface are also identified by X-ray photoelectron spectroscopy (XPS). These measurements show a significant increase in carbon every time the sample is placed under vacuum, while subsequent exposure to air causes an increase in oxidated carbon. Developing models for the contribution of carbon and oxygen to the membrane stress should therefore be an important future direction. Other contaminants, notably alkaline and halide ions, are detected in smaller quantities and briefly discussed.

## I. Introduction

Shift of resonance frequency in Microelectromechanical/Nanoelectromechanical systems (MEMS/NEMS) can be caused by either a quantifiable physical phenomenon to be measured, or unwanted instabilities that should be suppressed as much as possible. Sources of instabilities that affect the performance of MEMS can be classified as short-term or long-term effects that cause drift of certain properties over the course of many weeks. Thermomechanical fluctuations cause instant short-term instabilities. These fluctuations have been studied intensively in recent years in the context of high mechanical quality factor silicon nitride nanomechanical resonators [1]-[4]. Due to thermal energy, the atoms within the materials are randomly moving and colliding with each other. These excitations can couple to the resonator as a whole, creating mechanical fluctuations that often dictate the accuracy limit in short-term frequency measurements [5]. The mechanical properties of SiN enables it to be preloaded with a considerable amount of tensile stress, allowing it to reach high mechanical quality factors that isolate the device from such thermomechanical fluctuations. These low fluctuations make SiN resonators suitable, notably for acceleration [6], [7], pressure [8] and temperature sensing [9].

While short-term thermomechanical effects are now widely understood, instabilities from longer-term effects have not been as extensively studied. Such long-term instability will inevitably be relevant in eventual practical applications of SiN resonators, notably for calibration purposes. Long-term instabilities are expected to be caused by the aging of the device materials or by the slow adsorption/desorption of contaminants, which we aim to quantify in this work. Much of the knowledge on these aging phenomena have been developed in the context of quartz resonators in timing applications [10]-[12]. Due to their remarkably low drift and aging over a large period of time, quartz-based oscillators have been widely used in timing applications for several decades [13]. These devices showcase minimal aging effects leading to a very small frequency drift, often less than a part per million (ppm) per year [10]. Aging of quartz oscillators occur not only in the crystal itself but also in

M. Stephan, A. Bouchard, T. Hodges, and R. St-Gelais are with the Department of Mechanical Engineering, University of Ottawa, 75 Laurier Avenue East, Ottawa, Ontario K1N 6N5, Canada (email: mstephan@uottawa.ca; abouc067@uottawa.ca; thodg039@uottawa.ca; raphael.stgelais@uottawa.ca). T. Hodges, R. G. Green, T. Koukoulas and L. Wu are with the National Research Council Canada, 1200 Montreal Road, Ottawa, Ontario K1A 0R6, Canada (email: Richard.Green@nrc-cnrc.gc.ca; Triantafillos.Koukoulas@nrc-cnrc.gc.ca; Lixue.Wu@nrc-cnrc.gc.ca). R. St-Gelais is also affiliated with the Department of Physics, University of Ottawa, 75 Laurier Avenue East, Ottawa, Ontario K1N 6N5, Canada as well as the Nexus for Quantum Technologies Institute (NEXQT), University of Ottawa, 25 Templeton Street, Ottawa, Ontario K1N 6N5, Canada.

the electronic circuit paired with the crystal. In particular [10], quartz aging is found to be mainly due to contaminant adsorption/desorption, stress variation due to mount changes, material imperfections (dislocations, impurities, microcracks, surface and point defects), and diffusion of impurities from electrodes. These mechanisms [10] typically lead to a yearly stability of 0.5 to 2 ppm for TCXO (Temperature Compensation Crystal Oscillator) devices and 0.01 to 0.1 ppm for OCXO devices (Oven Controlled Crystal Oscillator). In more recent measurements [12], TCXOs and MCXOs (Microprocessor Compensation Crystal Oscillator) achieve yearly stabilities of respectively 0.5 and 0.02 ppm [12], while OCXOs are the most stable, achieving a yearly stability of 0.005 ppm [12].

Quartz crystal has been the default material for producing high-end oscillators for a long time, but in the past decades, the semiconductor industry led the development chip-integrated MEMS silicon-based clocks with performances on-par with quartz oscillators. Unlike standalone quartz, MEMS resonators are naturally integrated on silicon chips and are therefore simpler to integrate in semiconductor devices. These newer silicon timing devices, just like quartz oscillators, suffer from aging, notably induced by the adsorption/desorption of contaminants and stress variations in their mounting methods [14]. Using proprietary manufacturing and packaging processes (EpiSeal$^{TM}$ [15]), silicon MEMS produced commercially reach drift levels of 0.02 ppm/year, which is comparable to OXCO quartz oscillator devices [14].

While oscillator aging has been studied mostly in the context of timekeeping, relevant aging studies exist in other types of MEMS. Łuczak et al. [16] reports changes in the performance of two types of commercial silicon-based MEMS accelerometers that were studied for periods of 4 and 10 years. The authors studied both the offset (i.e., the output of a static device decoupled from gravity) as well as the scale factor (also commonly known as the sensitivity, which is the output voltage signal per external acceleration, in V/g) [16]. The accelerometer tested over 10 years resulted in average variations of 4100 ppm and 6780 ppm for the offset and scale factors, respectively [16]. Conversely, accelerometers tested over 4 years resulted in average variations of 1650 ppm and 4350 ppm [16].

Here we characterize aging of an as-fabricated low-stress SiN membrane over a period of 135 days. Motivation for this work is twofold. Firstly, developing a better understanding of the long-term frequency stability of SiN resonators will be of general interest for calibration purposes in practical applications of SiN resonant sensors. Secondly, a more direct question resulted from the development of mass-loaded SiN accelerometers [6] for seismic monitoring in the context of the Comprehensive Nuclear-Test-Ban Treaty Organization (CTBTO) [17]. For SiN devices to be acceptable in this context, their sensitivities (proportional to $\omega_0^{-2}$, where $\omega_0$ denotes the natural angular frequency) shall drift minimally, such that they remain traceable to internationally accepted primary standards [18].

## II. ADSORPTION/DESORPTION MODELS

While aging models for SiN are not readily available, previous work studied the effect of water adsorption/desorption on SiN cantilevers [19]. Since we expect significant water desorption from SiN in a vacuum environment, we review these models here and adapt them to our resonator parameters.

Adsorption of water on SiN resonators influences their eigenfrequency (*f*) through three sub-mechanisms. Firstly, the mass difference due to the adsorption/desorption of a contaminant ($\Delta m_{ads}$) such as water, will lead to small relative frequency ($\Delta f/f$) change by directly affecting the resonator mass ($m$) [5]:

$$\frac{\Delta f}{f} = -\frac{1}{2} \cdot \frac{\Delta m_{ads}}{m}. \quad (1)$$

For example, a partial monolayer adsorbed on a SiN resonator of surface area $A_{SiN}$ has a mass

$$m_{ads} = m_{mol} \cdot S_a \cdot A_{SiN} \cdot x, \quad (2)$$

where $m_{mol}$ is the mass of an adsorbed molecule (in kg), $S_a$ is the density of adsorption sites on the SiN surface, and $x$ is the fraction of occupied adsorption sites. In turn, $x$ can be calculated using the rates of adsorption ($r_a$) and desorption ($r_d$):

$$x = \frac{r_a}{r_a + r_d}. \quad (3)$$

These rates are respectively given by [1]

$$r_a = \frac{2}{5} \cdot \frac{P}{\sqrt{m_{mol} \cdot k_B \cdot T}} \cdot s \cdot A_{sites}$$

$$r_d = \nu_d \cdot e^{\left(-\frac{E_b}{k_B \cdot T}\right)}, \quad (4)$$

where $P$ is the partial pressure of the contaminant (1250 Pa for water in ambient setting and $\sim 10^{-5}$ Pa in high vacuum), $k_B$ is the Boltzmann constant ($1.38 \times 10^{-23}$ J/K), $T$ is the temperature of the resonator environment (297 K), $s$ is the sticking coefficient between the adsorbent and the adsorbates (0.1, taken from [1]), $A_{sites}$ is the area of the adsorption sites ($10^{-19}$ m$^2$, taken from [19]), $\nu_d$ is the desorption attempt frequency ($10^{13}$ Hz, taken from [1]) and $E_b$ is the desorption energy barrier ($1.1 \times 10^{-19}$ J, taken from [19]). Using these values, we find $x \approx 100\%$ at ambient pressure and $x \approx 0.16\%$ at a typical high vacuum pressure. We therefore expect the desorption of an almost complete monolayer during vacuum pump down. This corresponds to a $\Delta f/f = 600$ ppm frequency change caused by mass desorption for our specific resonator thickness ($h$ = 90 nm), while assuming one adsorption site per molecule at the SiN surface (i.e., $S_a = 5.7 \times 10^{18}$ sites/m$^2$).

Secondly, the surface tension of the water layer adsorbed on the SiN surface has its own built-in stress ($\Delta \sigma_{ST}$) that adds to the thin film stress of our device ($\sigma_{initial} \approx 70$ MPa), and thus affects its resonance frequency through

$$\frac{\Delta f}{f} = \frac{1}{2} \cdot \frac{\Delta \sigma_{ST}}{\sigma_{Initial}}. \quad (5)$$

In this case, we assume

$$\Delta \sigma_{ST} = \frac{\delta s}{h} x, \quad (6)$$

where $\delta s$ is the surface tension of the adsorbate (i.e., $\delta s = 0.073$ N/m for water, taken from [19]). Desorption of an almost complete monolayer of water during pump down (i.e., from $x$ = 100% to $x$ = 0.16 %) therefore yields a frequency change of -11.6%.

Thirdly, adsorbed water molecules will react chemically with silicon nitride, which induces an important source of additive chemisorption surface tension ($\delta s_{chem}$) in a similar fashion as equations (5) - (6), i.e., $\Delta\sigma_{chem} = \delta s_{chem} x/h$. In [19], this chemisorption surface is estimated as $\delta s_{chem}$= 1.1 N/m, leading to the strongest frequency shift (-17.5%) of the three effects, for the same monolayer desorption.

Obviously, in an experimental setting where we do not immediately track frequency at the beginning of a pump down, we expect to see only the tail end of the effect of water desorption (i.e., much less than the predicted drift). Nevertheless, these models are qualitatively instructive for two main reasons. Firstly, they indicate that drift from water desorption should be negative (i.e., the mass desorption contribution is negligible compared to surface tension-based stresses). Secondly, it should be noted that the calculated molecule desorption rate ($r_d = 23 \text{ s}^{-1}$) is far from being instantaneous, compared for example, with adsorption at ambient pressure ($r_a = 4.5 \times 10^5 \text{ s}^{-1}$). We can therefore reasonably expect water desorption to be noticeable over a relatively long time, especially if water molecules occupy deep trap imperfections (i.e., with higher $E_b$) at the resonator surface.

## III. Experimental Methods

We characterize a 6.45 × 6.45 mm low-stress (70 MPa) SiN membrane with a thickness of 90 nm, fabricated using the process given in [20]. The membrane is secured on an in-house engineered mount, shown in Fig 1 (b), that uses spherical magnets to hold our device in place, thus minimizing lateral mounting stresses. A low-carbon steel magnetic plate with purposely designed divots is used to host the lower layer of magnets comprised of three 1 mm spherical N52 neodymium balls with a pull force of 0.5 lb. This magnetic mounting technique minimizes mounting stress and avoids fast aging materials (e.g., adhesives) that would interfere with the aging study of the SiN nanomechanical resonator.

The mount assembly hosts a single mode optical fiber that is positioned to point towards the membrane. Fig 1 (d) schematizes the optical interferometry system, mainly comprised of a 90:10 coupler, a Thorlabs PDA20CS2 photodetector, and a JDS Uniphase SWS15101 tunable laser outputting 5 mW at a wavelength of 1540 nm, that is used to measure the membrane resonance frequency [21]. The frequency of the fundamental eigenmode as well as degenerate mode 1-2 is constantly tracked throughout the aging measurements using a digital phase-locked loop (PLL). The PLL module is provided by a Zurich Instruments MFLI Lock-in Amplifier that is under valid calibration and that has a TCXO reference oscillator with a rated long-term stability of ±1 ppm/year [22].

Several control tests were conducted to confirm that the laser used for measuring the membrane resonance frequency has negligible interference with our aging measurement. A first control test, showcased in supplementary Fig S1, was conducted to study the potential effects of membrane heating caused by the readout laser. A variable attenuator is actuated from 0 to 5 V, which respectively attenuates the laser power from 1.5 down to 25 dB. This high magnitude attenuation resulted in less than 300 ppm of frequency change. Since the laser power is rated to fluctuate by less than 0.01 dB per hour, we expect less than 0.1 ppm drift due to laser heating. We nevertheless added a 3 dB attenuator to the optical system as extra caution. Likewise, in supplementary Fig S2, we characterize the potential effect of changes to the interference condition (i.e., between constructive and destructive) as the membrane-optical fiber distance drifts. Changing the wavelength of the laser in steps of 0.015 nm allows scanning of the whole interference fringe and results in less than 8 ppm frequency changes.

To suppress air damping on our resonator and maximize its quality factor, the mount assembly is placed inside a vacuum cell, shown in Fig 1 (a), where the vacuum is first initiated by an Edwards T-Station 85 turbomolecular roughing pump to then be maintained around $10^{-8}$ Torr using a Gamma Vacuum 3S Titan DI noble diode ion pump.

SiN membrane resonators are extremely sensitive to temperature changes in their environment, such that temperature drift must be measured and compensated for to obtain the net aging. To minimize temperature drifts, the vacuum cell containing the membrane is placed inside a Measurement International 9300 air bath, pictured in Fig 1 (d) which provides a stable thermal environment with variations of ±50 mK for every ± 1°C temperature difference with the ambient environment. Additionally, a calibrated Keithley DMM6500 multimeter is used to continuously gather the four-wire resistance of a Measurement Specialties 55036 Glass NTC calibrated thermistor located inside the vacuum cell. This cell temperature (also denoted $T_{cell}$ in Fig 2) measurement is then used for compensating the membrane frequency measurements and eliminating the frequency drift signal linked to temperature changes. The multimeter also gathers the temperature of the ambient environment by measuring the two-wire resistance of a TE Connectivity Measurement Specialties NB-PTCO-186 resistance temperature detector (RTD). This measurement is denoted as "Lab Temperature" in Fig 2 and after. The recorded temperature and frequency signals are continuously logged for 135.75 days (limited by instrument availability) using a designed Python script that stores the obtained data and saves it in structured text files.

After the aging measurements, X-ray photoelectron spectroscopy (XPS) is used to characterize surface contamination present on the aged sample as well as two control samples. XPS is sensitive to elemental composition found in the top 5 nm – 7 nm of the sample surface but is insensitive to hydrogen and helium. Spectra are acquired with a Kratos Axis Ultra operating at a nominal base pressure of 4 x 10$^{-9}$ Torr, using a monochromated aluminum X-rays source (Kα 1486.7 eV), with the charge neutralizer enabled. The instrument binding energy scale is calibrated using metallic gold and copper samples to yield a binding energy (BE) of 83.96 eV and 932.62 eV for the Au 4f7/2 and Cu 2p3/2

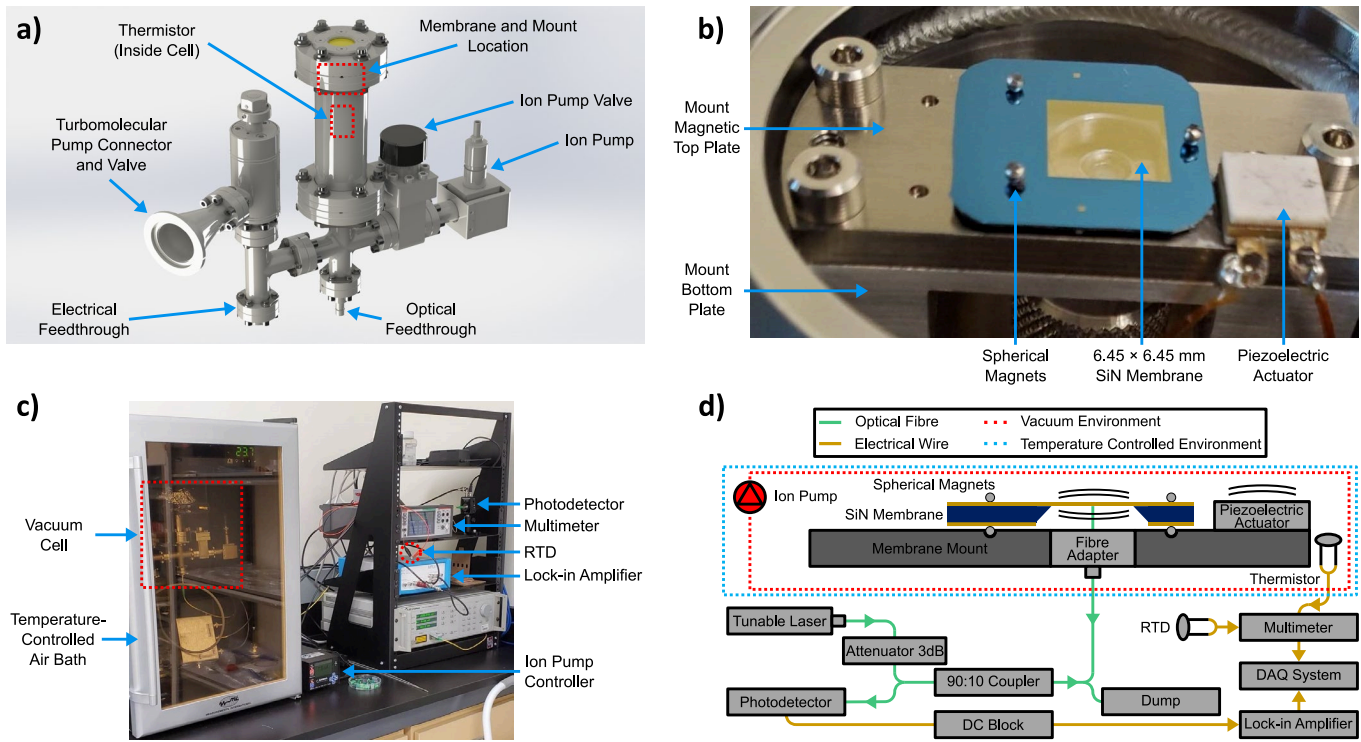

Fig. 1 (a). Vacuum cell used to hold the membrane resonator under high vacuum during aging measurements, (b) SiN membrane used in the aging measurement, magnetically clamped on a custom-designed holder (c) Picture of the different instruments and devices used for aging measurements. The vacuum cell of panel (a) is placed inside a temperature-controlled air bath. (d) Functional schematic of the measurement setup.

transitions, respectively. The incident X-ray beam produces a photoemission current of 1.2 nA, from which an area of 700 µm × 300 µm is analyzed. Measurement spectra are acquired from the underside (fiber detection side) of the silicon nitride membrane at 3 to 4 different locations distributed diagonally along its surface such that there is no overlap in the incident X-ray beam between locations. Broad region survey spectra (80 eV pass energy, 1 eV step, 1 s dwell) and higher resolution spectra (20 eV, 0.1 eV step, 2 s dwell) of the main transitions studied are acquired at each position. At one of the positions, angle-resolved (ARXPS) survey spectra are acquired. Tilting the sample surface normal with respect to the analyzer lens axis increases surface sensitivity by increasing the path length of electrons through the sample surface. Thus, providing information regarding the relative depth of elements at the surface. Spectra are analyzed using CasaXPS software version 2.3.17 [23], with the binding energies referenced to the main peak (C-C/C-H) of the carbon 1s spectrum of adventitious carbon set to 284.8 eV [24]. Tougaard backgrounds are used to fit the main components of the spectra [25], while linear backgrounds are used for the contaminants. The spectra intensities are corrected using empirical Relative Sensitivity Factors (RSFs) from the Kratos element library and for the instrument transmission function. Unless indicated, the quantification results are not corrected for the carbon overlayer, which will result in a slight overestimation of transitions involving electrons of higher kinetic energies (lower binding energies). The thickness of the carbonaceous overlayer and correction of the atomic concentrations due to the escape depth through the carbon layer are estimated using the method presented by Smith [26]. The thickness calculated using this method is converted to mass through the surface area of the membrane, and the density of the overlayer is assumed to be 2.1 ± 0.5 g/cm$^3$, though variations have been reported in the literature depending on the nature of the overlayer and the hybridization of carbon. The latter is attributed largely to sp3 diamond-like carbon 1.09 g/cm$^3$ to 3.32 g/cm$^3$ [27]. Spectra from each position are averaged for the determination of chemical atomic concentrations, while the standard deviation between positions is reported as the uncertainty on the reported values.

## IV.  RESULTS

Raw data signals for aging measurement (shown in Fig 2 (a)) include the membrane resonance frequency for eigenmodes 1-1 (in orange) and 1-2 (in blue), and temperature in the vacuum chamber (in green). Frequency data is streamed at a rate of 7 samples/s, while temperature data is streamed at a rate of 30 samples/s. In this raw signal, we note that eigenmode 1-1 is much more stable than mode 1-2. It is likely that the degeneracy of mode 1-2 leads to mixing with mode 2-1 in a fashion that varies over time, thus increasing drift. The same possibility does not exist for non-degenerate mode 1-1. From the beginning of the measurements, the eigenfrequency of mode 1-2 continuously drifts in an upward fashion with an overall magnitude of 60 Hz. On the other hand, the frequency of mode 1-1 is more stable, fluctuating by approximately 10 Hz around a stable central value of 18,210 Hz.

The frequency of mode 1-1 follows closely the temperature changes in the vacuum cell ($T_{cell}$). A cyclic behaviour is observed for both the chamber temperature and the frequency of mode 1-1, whereas measurements alternate between a lower and higher plateau of values throughout the entire measurement period. This cyclical temperature change is linked to the heating and cooling of our laboratory environment. Fig 2 (b) includes the lab temperature measured every 5 minutes by a RTD, confirming that the thermal variations in the vacuum cell (with an amplitude of about 0.3 K) are strongly correlated to the much larger thermal variations in the laboratory, on the order of 2 K.

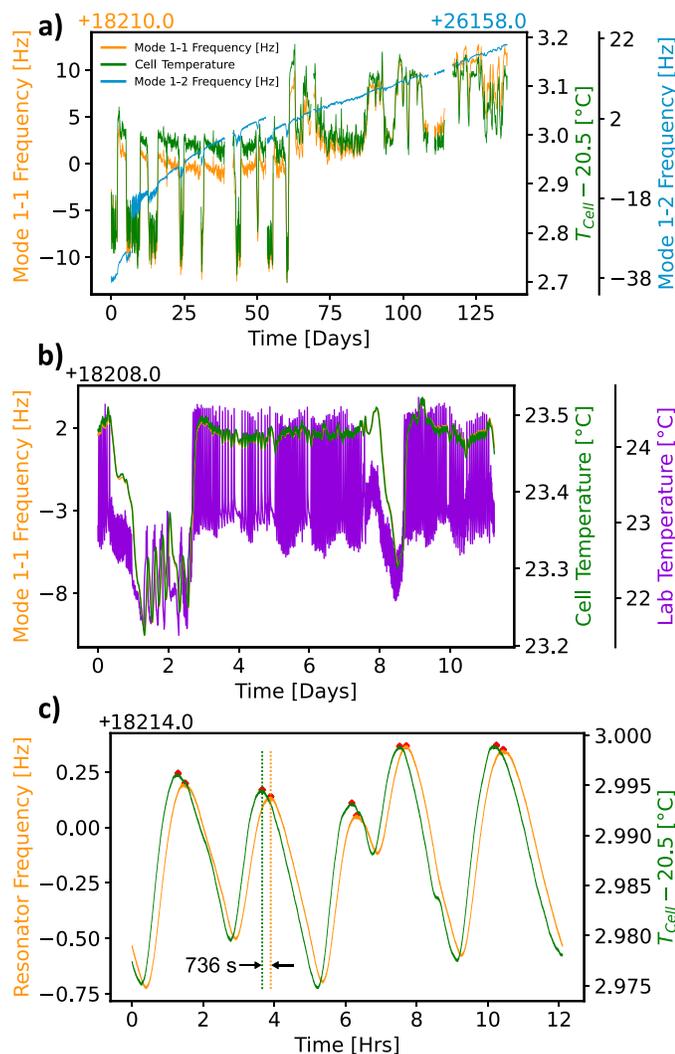

Fig. 2 (a). Raw measured frequency signals for membrane eigenmode 1-1 (orange) and mode 1-2 (blue), together with temperature of the vacuum cell (green), throughout the aging study. (b) The measured temperature of the vacuum cell (green) and of mode 1-1 frequency (orange) are found to be strongly correlated to the measured temperature of the lab environment (purple). (c) Example of thermal time lag between the vacuum cell temperature readings and the corresponding frequency shift of membrane eigenmode 1-1.

To extract the net membrane eigenfrequency drift, temperature compensation of the frequency measurements is performed using the measured cell temperature. To do so, we must first correct for a thermal lag time between the measured cell temperature and the membrane temperature. This lag originates from the large thermal mass of the membrane holder compared to the $T_{cell}$ measurement thermistor that is suspended in the middle of the chamber at the location indicated in Fig 1 (a) (see also supplementary Fig S3). As shown in Fig 2 (c), this lag averages around 736 seconds and is subtracted from the data using a custom peak finder script.

With the thermal lag suppressed, the linear dependence between the membrane resonance frequency and the temperature of the vacuum cell becomes evident. In Fig 3 (a), a colormap presents the frequency variations of the membrane over time and as a function of the lag-compensated cell temperature. In this representation, a membrane suffering from no drift other than thermal would yield a single diagonal line. In Fig 3 (a), such a trend is clearly visible, but the vertical position of the linear curve slowly changes over time. This change is the temperature-compensated aging that we wish to measure. To do so, we perform a linear fit of the data of Fig 3 (a) every 6 hours to produce the final temperature-compensated aging plot in Fig 3 (b). These fits also provide the sensitivity of the membrane eigenfrequency to temperature, i.e., 46.8 Hz/K or 2570 ppm/K measured here for mode 1-1.

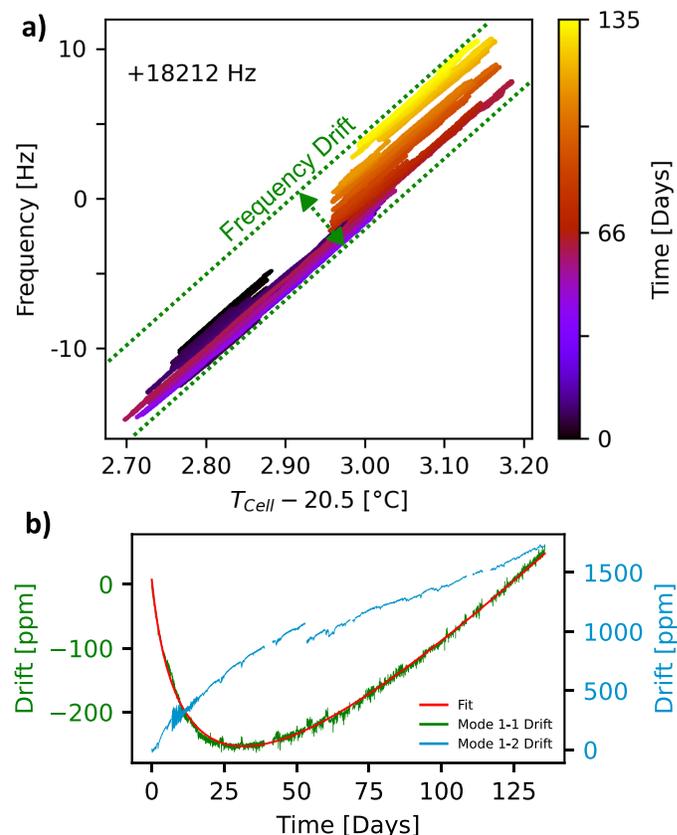

Fig. 3 (a). Color plot depicting the relationship between the frequency of eigenmode 1-1 and the temperature of the cell over time. A membrane suffering from no drift other than thermal would have points along a single diagonal line. (b) Final temperature-compensated stability plot throughout the aging study. A double logarithmic trend (in red) fits the data well, as is also the case in quartz resonator aging measurements.

The last temperature-compensated aging drift plot in Fig 3 (b), quantifies the stability of the fundamental resonance mode (in green) and mode 1-2 (in blue). The stability of mode 1-1 (300 ppm for 135 days of aging) is significantly higher than that of mode 1-2 (1750 ppm for 135 days of aging). As discussed above, we hypothesize that higher drift for mode 1-2 is a consequence of varying mode hybridization with mode 2-1. We therefore focus our discussion on mode 1-1. Interestingly, a double logarithmic trend line (fitted in Fig 3b) is found to match well with the aging data for mode 1-1, as is also typically the case in quartz resonators aging measurements [10].

The measured frequency drift of 0.03% ppm is much smaller than the overall drift of 18.6% predicted for desorption of water in section II. This discrepancy is not surprising since the frequency data is not recorded during the initial hours of the vacuum pump down, i.e., at a critical period where desorption of water, and possibly of other volatile compounds, occurs. Data acquisition began 48 hours after pump down, a delay necessary to allow turbomolecular pre-pumping (24 hours) before switching to the ion pump and stabilization of the cell temperature when moved in the temperature-controlled bath (24 hours).

The aging of our resonator (300 ppm over 135 days) is much larger than the values measured for most quartz TCXO (0.5 ppm over 365 days [12]). However, when normalized to the thickness of the devices, around 90 nm for SiN and 100 μm for TCXOs [28], aging values are in the same order of magnitude: i.e., 70 ppm · μm · year$^{-1}$ for SiN and 50 ppm · μm · year$^{-1}$ for TCXOs [12]. This strengthens the commonly accepted assumption that aging is largely a surface effect [10] and that good resonators for timekeeping should have large thicknesses, as opposed to SiN membrane resonators.

After around 30 days of aging data, we observe a reversal in aging direction, and measured aging then follows a positive trend. At this stage, it is possible that water desorption becomes negligible and that aging becomes linked to adsorption and/or desorption of other contaminants. In an effort to identify possible contaminants, XPS measurements are therefore performed on the aged membrane as well as two other control samples that were fabricated at the same time as the aged device.

To characterize the surface composition of the vacuum aged sample, we perform X-ray photoelectron spectroscopy (XPS) measurements on the aged membrane and on the two control membranes that were stored in laboratory air in a plastic (typically polystyrene or polycarbonate [29]) laboratory grade Petri dish. All membranes share the same architecture as well as geometry and were manufactured in the same batch of nanofabrication that ended with KOH etching followed by deionized water rinsing. The aged membrane characterized throughout this aging study is labeled as AM1. The first control membrane (CM1) was measured at the same time as AM1 throughout the XPS studies. The second reference membrane (CM2) was introduced later as an extra control. Example scans fore these three samples are given in Fig 4 (a).

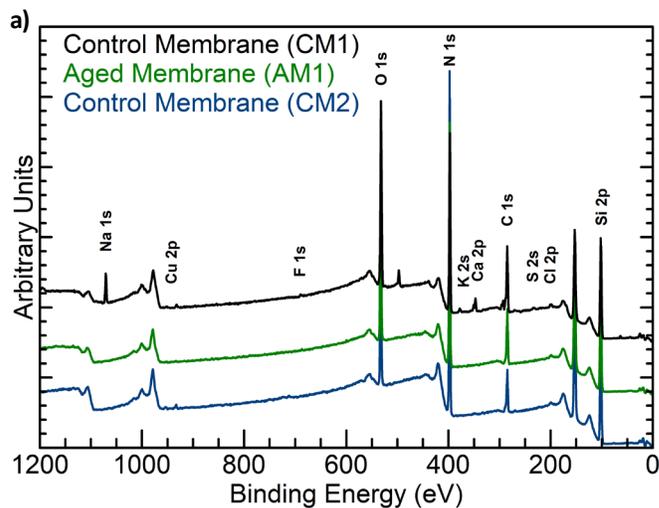

a)

b)

| Transition | Position Averaged Concentration (atm.%) | | |
|---|---|---|---|
| | AM1 | CM1 | CM2 |
| Si 2p | 33.3 ± 0.2 | 31.7 ± 4.2 | 34.9 ± 0.4 |
| N 1s | 34.4 ± 0.1 | 34.4 ± 5.7 | 38.7 ± 0.8 |
| O 1s | 16.2 ± 0.2 | 17.9 ± 2.2 | 15.6 ± 0.7 |
| C 1s | 15.7 ± 0.4 | 13.9 ± 5.4 | 10.4 ± 0.2 |
| Na 1s | ND | 0.66 ± 0.62 | ND |
| Cu 2p | 0.03 ± 0.01 | 0.05 ± 0.02 | 0.09 ± 0.02 |
| F 1s | 0.05 ± 0.02 | 0.06 ± 0.10 | 0.04 ± 0.04 |
| Ca 2p | ND | 0.34 ± 0.34 | ND |
| K 2p | ND | 0.34 ± 0.28 | ND |
| S 2s | ND | 0.30 ± 0.27 | ND |
| Cl 2p / Si 2s Loss | 0.05 ± 0.02 | 0.34 ± 0.10 | 0.04 ± 0.04 |

Fig. 4 (a). X-ray photoelectron spectroscopy (XPS) plot (single-position example scan) for each membrane upon first entry into analysis chamber. Presence of surface contaminants for control membrane 1 (CM1) is significantly higher than the aged membrane (AM1). (b) XPS table indicating the concentration of surface contaminants for control membranes and the aged membrane. The concentrations are averaged from measurements performed at three different locations on the membranes. ND: Not detected, the peak area is lower than the fit uncertainty and repeatability for all positions measured for the membrane.

Survey spectra of the of the vacuum aged and air stored samples showcase similar compositions upon first introduction into the analysis chamber (see Fig 4 b). These spectra demonstrate a level of carbon contamination from adventitious carbon (AdC) [30] of approximately 10 – 15 atomic % (atm.%), which corresponds to an estimated overlayer thickness of 0.4 – 0.6 nm. Some samples have additional trace levels of contamination, of less than 1 atm %. The major contaminants found are the alkaline elements calcium (Ca), potassium (K) and sodium (Na) as well as halide chlorine (Cl). Fluorine (F) is found in trace amounts on samples CM1 and AM1 but not on CM2, copper (Cu) is present in small amounts (< 0.1 atm).

Copper is a well known contaminant in silicon and is difficult to remove [31], [32]. The presence of oxygen is attributed to the formation of silicon dioxide or silicon oxynitride which is known to form as a native oxide on the surface of silicon nitride in air [33]. Additionally, oxygen can originate from chemisorbed water and oxygen bearing carbon species that are typically found in adventitious carbon [30]. Surface carbon is found on all samples when first measured by XPS. Potassium is conclusively resolved on all positions of sample CM1, but its presence is not conclusively observed on all other membranes. Finally, we note that even though Chlorine (Cl 2p) peaks are noticeable on all samples, we cannot rule out the possibility of interference from Si 2s loss occurring at the same biding energy, hence the Cl 2p / Si 2s notation if Fig 4 (b).

The initial XPS measurements comparing AM1 to CM1, as presented in Fig 4, indicate a lower concentration of contaminants for the aged sample compared to the first control membrane. This could be explained by desorption due to the long period spent under vacuum for AM1. This hypothesis is, however, refuted by XPS measurements on an additional control membrane CM2 which revealed a lesser contaminant concentration than AM1 despite having not spent any time under vacuum and being stored in the same laboratory environment as CM1. Therefore, contaminant concentration (other than for the case of carbon discussed below) appears to be mainly impacted by variations in manufacturing and handling, rather than by time spent in ambient air.

The concentration of carbon on the sample surface is found to significantly increase each time the sample was extracted from the XPS analysis chamber for storage in air and then returned to the XPS chamber for analysis under vacuum. The observed increase does not correlate with the time spent in air; rather, it is found to correlate strongly with the number of air-vacuum cycles the sample is exposed to. Fig 5 demonstrates this correlation, with samples CM1 and AM1 increasing carbon concentration by an average of 3% per air-vacuum cycle. Mass increase due to carbon uptake with vacuum cycles has been reported previously for mass standards [35]. Angle XPS resolved measurements (Supplementary Fig S4) show that carbon (and oxidated carbon) uptake occurs primarily on the surface of SiN rather than in the bulk of the material. It is possible that carbon adsorption causes the frequency increase, starting at 30 days in Fig 3 (b), if it causes positive chemisorption stress on SiN as is the case for water. Chemisorption stress for carbon is known to occur on metals [36]. It would be interesting to study its effect on SiN in light of these results.

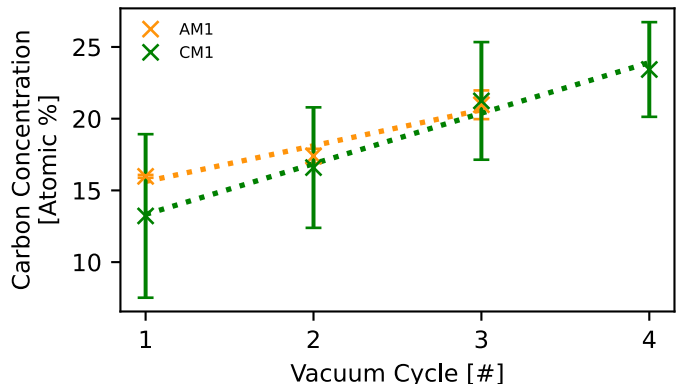

Fig. 5. Change in the carbon concentration with each air-vacuum cycle that sample AM1 (orange) and CM1 sustained (green).

## V. Conclusion

We presented measurements and analysis of the long-term stability of as-fabricated silicon nitride (SiN) nanomechanical resonator devices. Our measurements provide a baseline value for samples stored in a typical lab environment without specific treatment to improve cleanliness. Improvements on these measurements are likely possible, for example, using thermal or chemical passivation and/or cleaning treatments immediately before usage under vacuum. Our developed measurement and temperature compensation protocol will provide useful methods for measuring these improvements in the future. Interestingly, the aging trend observed for our resonator is very similar to the double logarithmic models found in literature for other types of oscillators, such as quartz. Likewise, the drift amount due to aging of our membrane is comparable to that of quartz when weighing the drift value to the resonator thicknesses. This strengthens the commonly accepted assumption that aging of micro/nano resonators is essentially a surface (rather than bulk) phenomenon. We identified water and carbon as two highly probable adsorbates causing these surface effects. However, as is the case for aging studies on quartz [10], their exact contribution during our measurement is not individually measurable in real-time and therefore remains debatable. For example, it is possible that the initial negative aging trend is caused by residual water desorption, followed by positive aging due to carbon adsorption in the vacuum environment. In this hypothesis, the desorption of water liberates adsorption sites for the subsequent adsorption of hydrocarbons. It is also possible that water is completely desorbed by the time we begin our frequency tracking measurements, and that the observed aging is due purely to carbon contamination. We note that there exist reports for laser [37] and X-ray [38] stimulated adsorption of hydrocarbons in a vacuum environment. This is usually observed for higher intensity sources than used for this study, however given the sensitivity and the timescales considered, this mechanism may be contributing. In any case, the model provided here will be useful for considering the effect of water desorption in any future measurements on SiN, and indicates that modeling chemisorption stress for other species, such as carbon, is required.

# SUPPLEMENTARY

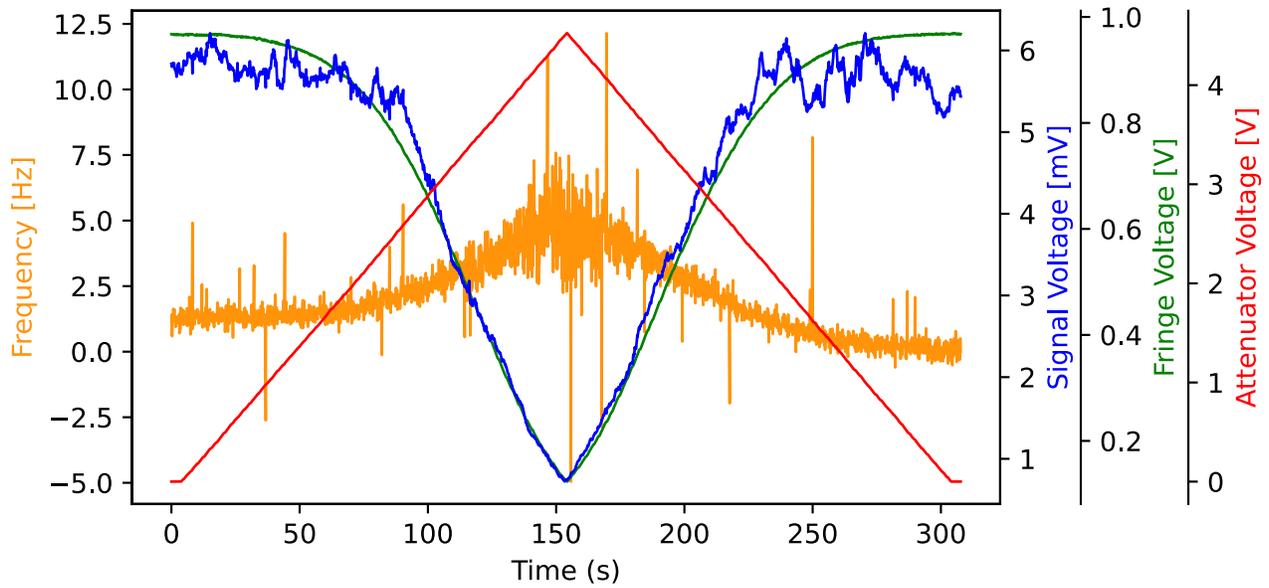

Fig S1. Laser heating test that depicts the change in frequency the membrane undergoes as a function of laser power, which is controlled by a variable attenuator actuated between 0 and 5V to respectively. Attenuating the laser power from 1.5 to 25 dB results in a $\approx 5$ Hz (270 ppm) frequency change, such that we expect less 0.1 ppm drift considering the frequency stability of our laser.

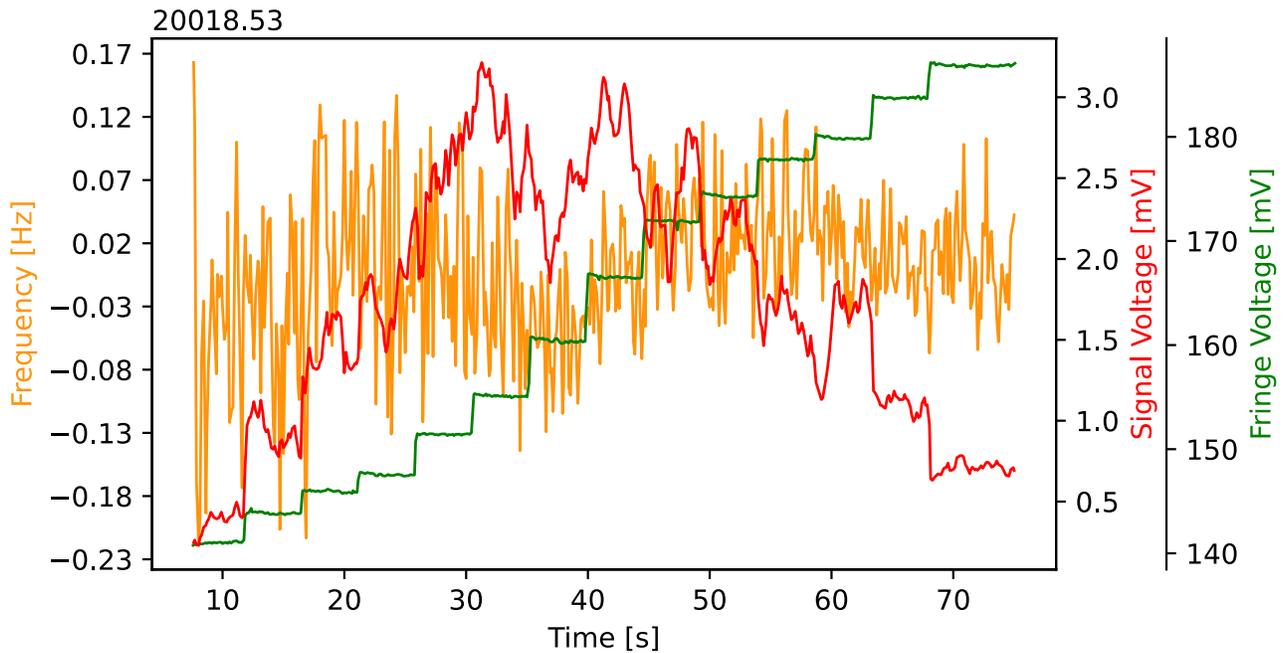

Fig S2. Wavelength variation test that depicts the change in frequency of the membrane as a function of the laser interference condition (i.e., the "fringe voltage" going from constructive to destructive interference). A maximum 0.25 Hz drift (<15 ppm) is therefore expected if the fiber-membrane distance drifts by a complete half wavelength.

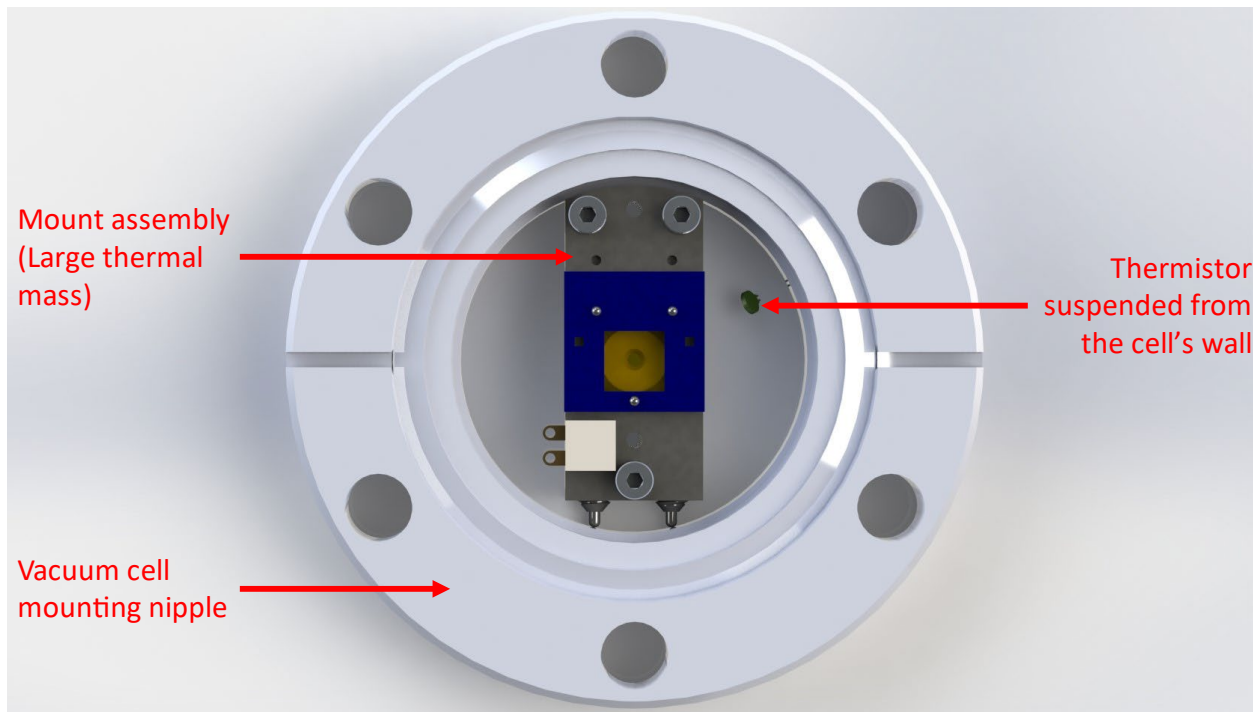

Fig S3. CAD rendering of the membrane assembled on the designed mount inside the vacuum cell, alongside the thermistor hanging from the wall of the cell.

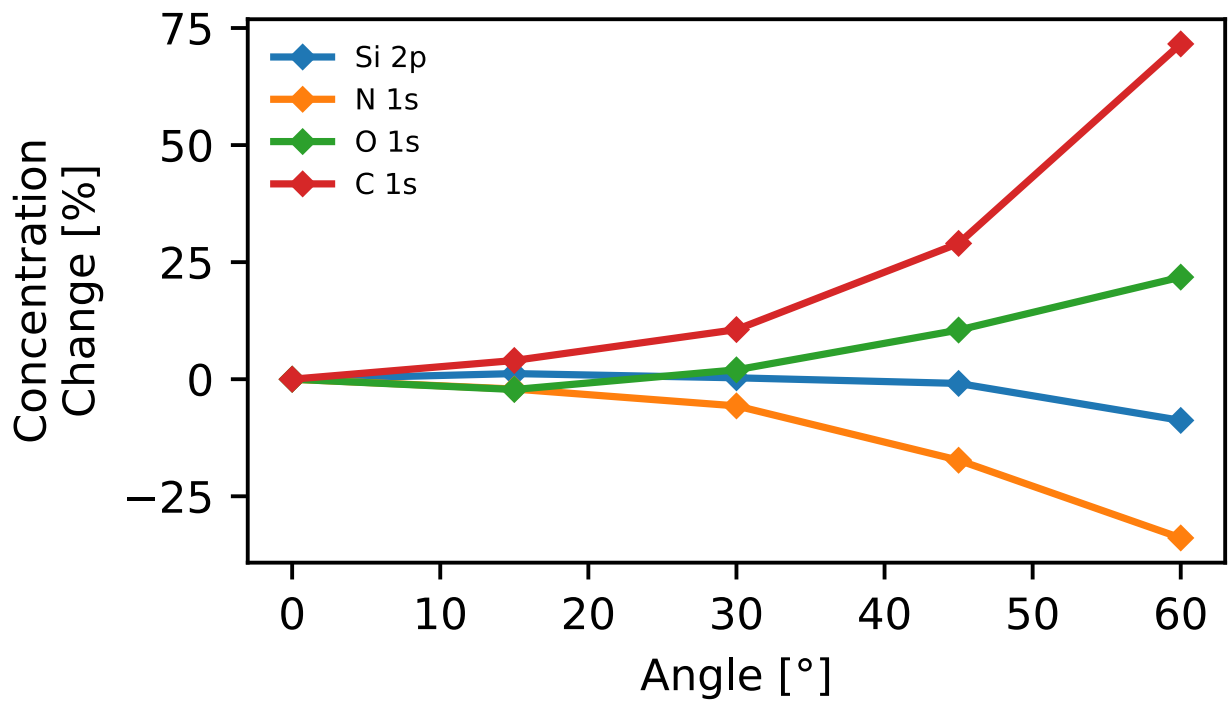

Fig S4. Survey scans of each membrane measured upon the first entry of each sample into the analysis chamber, performed at four additional angles from the surface normal. This can be used to determine the relative depth of elements in the surface via the change in their apparent concentration as the angle is increased and the measurements become increasingly surface sensitive. The observed enhancement factor, given as the change in concentration with angle relative to the value at 0° (the sample surface normal to the analyzer axis), is presented for sample CM1, though all samples exhibit similar trends. The largest signal enhancement is seen for carbon, indicating it lies closest to the surface. This is followed by oxygen which shows a moderate increase. Silicon and nitrogen signals both decrease, with nitrogen presenting the largest decrease, confirming it lies at the largest depth.